\documentclass[epj]{webofc}
\usepackage[varg]{txfonts}   
\usepackage{hyperref}

\renewcommand*{\eqref}[1]{Eq.~(\ref{eq:#1})}

\newcommand*{\figref}[1]{Fig.~(\ref{fig:#1})}
\newcommand*{\figlab}[1]{\label{fig:#1}}

\newcommand*{\seclab}[1]{\label{sec:#1}}
\wocname{ARENA-2016}
\woctitle{ARENA-2016}
\begin{document}
\title{The Giant Radio Array for Neutrino Detection}
%
%

\author{The GRAND collaboration:~\firstname{Olivier} \lastname{Martineau-Huynh}\inst{1}\fnsep\thanks{\email{omartino@in2p3.fr}}
\and \firstname{Mauricio} \lastname{Bustamante}\inst{2} 
\and \firstname{Washington} \lastname{Carvalho}\inst{3} 
\and \firstname{Didier} \lastname{Charrier}\inst{4} 
\and \firstname{Sijbrand} \lastname{De Jong}\inst{5} 
\and \firstname{Krijn D.} \lastname{de Vries}\inst{6}
\and \firstname{Ke} \lastname{Fang}\inst{7}
\and \firstname{Zhaoyang} \lastname{Feng}\inst{8}
\and \firstname{Chad} \lastname{Finley}\inst{9}
\and \firstname{Quanbu} \lastname{Gou}\inst{8}
\and \firstname{Junhua} \lastname{Gu}\inst{10}
\and \firstname{Hongbo} \lastname{Hu}\inst{8}
\and \firstname{Kumiko} \lastname{Kotera}\inst{11}
\and \firstname{Sandra} \lastname{Le Coz}\inst{10}
\and \firstname{Clementina} \lastname{Medina}\inst{1,12}
\and \firstname{Kohta} \lastname{Murase}\inst{13}
\and \firstname{Valentin} \lastname{Niess}\inst{14}
\and \firstname{Foteini} \lastname{Oikonomou}\inst{13}
\and \firstname{Charles} \lastname{Timmermans}\inst{5}
\and \firstname{Zhen} \lastname{Wang}\inst{8}
\and \firstname{Xiangping} \lastname{Wu}\inst{10}
\and \firstname{Yi} \lastname{Zhang}\inst{8}
}


\institute{LPNHE, CNRS-IN2P3 and Universit\'es Paris 6 \& 7, BP200, 4 place Jussieu, 75252 Paris, France
\and
Center for Cosmology and AstroParticle Physics, The Ohio State University, Columbus, OH 43210, USA
\and
Physics institute, University of S\~ao Paulo, Rua do Mat\~ao, trav. R, Cid. Universit\'aria, S\~ao Paulo, Brazil
\and
SUBATECH, CNRS-IN2P3, Universit\'e de Nantes, Ecole des Mines de Nantes, Nantes, France
\and
Nikhef/Radboud University, Nijmegen, the Netherlands
\and
Vrije Universiteit Brussel, Dienst ELEM, B-1050 Brussels, Belgium
\and
Department of Astronomy, University of Maryland, College Park, MD, 20742
\and
Key Lab of Particle Astrophysics, IHEP, Chinese Academy of Sciences, Beijing 100049, China
\and
Oskar Klein Centre and Dept. of Physics, Stockholm University, SE-10691 Stockholm, Sweden
\and
National Astronomical Observatory, Chinese Academy of Sciences, Beijing 100012, China
\and
IAP, Sorbonne Universit\'es, Paris 6 and CNRS, 98 bis bd Arago, 75014 Paris, France
\and
Instituto Argentino de Radioastronomia, CCT La Plata-CONICET, 1894, Villa Elisa C.C. No. 5, Argentina
\and
Dept. of Physics, Dept. of Astronomy \& Astrophysics, Penn State University, University Park, PA, USA
\and
LPC, CNRS-IN2P3, Universit\'e Blaise Pascal, BP 10448, 63000 Clermond-Ferrand, France
}

\abstract{%
The Giant Radio Array for Neutrino Detection (GRAND) is a planned array of $\sim$ 2$\cdot10^5$ radio antennas deployed 
over $\sim$\,200\,000\,km$^2$ in a mountainous site. It aims primarly at detecting high-energy neutrinos via the observation of extensive air showers induced by the decay in the atmosphere of taus produced by the interaction of cosmic neutrinos under the Earth surface. GRAND aims at reaching a neutrino sensitivity of $5\cdot10^{-11}E^{-2}$\,GeV$^{-1}$ \,cm$^{-2}$\,s$^{-1}$\,sr$^{-1}$ above $3\cdot10^{16}$\,eV. This ensures the detection of cosmogenic neutrinos in the most pessimistic source models, and $\sim$50 events per year are expected for the standard models. The instrument will also detect UHECRs and possibly FRBs. Here we show how our preliminary design should enable us to reach our sensitivity goals, and discuss the steps to be taken to achieve GRAND, while the compelling science case for GRAND is discussed in more details in \cite{ICRC2015}.
}
\maketitle
%


\section{Detection Method}\seclab{method}

Cosmic $\nu_{\tau}$ can produce taus\footnote{Other neutrino flavors can be neglected, as the electron range in matter at these energies is too short and the muon decay length too large.} under the Earth surface through charged-current interactions. Taus may then exit and decay in the atmosphere, generating Earth-skimming extensive air showers 
(EAS)~\cite{Fargion00,Bertou04}. EAS emit coherent electromagnetic radiations at frequencies of a few to hundreds of MHz, detectable by radio antennas for shower energies $E \gtrsim 3\cdot10^{16}$ eV~\cite{CODALEMA2005,LOPES2005}. 
The strong beaming of the electromagnetic emission, combined with the transparency of the atmosphere to radio waves, will allow the radiodetection of EAS initiated by tau decays at distances up to several tens of kilometers (see \figref{fig}), making radio antennas ideal instruments for the search of cosmic neutrinos. Furthermore, antennas offer practical advantages (e.g. limited unit cost, easiness of deployment) that allow the deployment of an array over very large areas, as required by the expected low neutrino event rate. 

Remote sites, with low electromagnetic background, should obviously be considered for the array location. In addition, mountain ranges are preferred, first because they offer an additional target for the neutrinos, and also because mountain slopes are better suited to the detection of Earth-skimming showers compared to flat areas which are parallel to the neutrino-induced EAS trajectories.

GRAND antennas are foreseen to operate in the $30-100$\,MHz band. Below this range, short-wave background prevents detection, while coherence of radio emission fades above it.
However, an extension of the antenna response up to 200 or 300\,MHz would enable us to better observe the Cherenkov ring associated with the air shower \cite{Alvarez2012}, which represents a sizable fraction of the total electromagnetic signal at these frequencies. This could provide an unambiguous signature for background rejection.

\section{GRAND layout and neutrino sensitivity}\seclab{sensitivity}
We present here a preliminary evaluation of the potential of GRAND for the detection of cosmic neutrinos, based on the simulated response of a 90\,000 antennas setup deployed on a square layout of 60\,000\,km$^2$ in a remote mountainous area, the Tianshan mountains in the XinJiang province, China. 

{\bf Simulation method.}
We perform a 1D tracking of a primary $\nu_{\tau}$, simulated down to the converted tau decay. 
We assume standard rock with a density of 2.65~g/cm$^3$ at sea level and above, while the Earth core is modeled following the Preliminary Reference Earth Model \cite{Dziewonski81}. 
The simulation of the deep inelastic scattering of the neutrinos is performed with Pythia6.4, using the CTEQ5d probability distribution functions (PDF) combined with \cite{Gandhi98} for cross section calculations. The propagation of the outgoing tau is simulated using randomized values from parameterisations of GEANT4.9 PDFs for 
tau path length and proper time. Photonuclear interactions in GEANT4.9 have been extended above PeV energies following \cite{Dutta00}. The tau decay is simulated using the TAUOLA package. The radiodetection of neutrino-initiated EAS is simulated in the following way:~ 

- for a limited set of $\nu_{\tau}$ showers simulated with ZHaireS \cite{Zhaires} at various energies (see \figref{fig}), we determine a conical volume inside which the electric field is above the expected detection threshold of the GRAND antennas (30~$\mu$V/m in an agressive scenario, 100~$\mu$V/m in a conservative one). 

- from this set of simulations, we parametrize the shape (angle at top and height) of this detection cone as a function of energy. 

- for each neutrino-initiated EAS in our simulation, we compute the expected cone shape and position, and select the antennas located inside the corresponding volume, taking into account signal shadowing by mountains. 

- if a cluster of 8 neighbouring units can be found among these selected antennas, we consider that the primary $\nu_{\tau}$ is detected. 

{\bf Results and implications.}
Assuming a 3-year observation with no neutrino candidate on this 60\,000 km$^2$ simulated array, a 90\%\,C.L. integral limit of $6.6\times10^{-10}$~GeV$^{-1}$~cm$^{-2}$~s$^{-1}$ can be derived for an $E ^{-2}$ neutrino flux in our agressive scenario~($1.3\times10^{-9}$ in our conservative scenario). This is a factor $\ge 5$ better than than other projected giant neutrino telescopes for EeV energies \cite{ARA2016}.

This preliminary analysis also shows that mountains constitute a sizable target for neutrinos, with $\sim$50\% of down-going events coming from neutrinos interacting inside the mountains. 
It also appears that specific parts of the array (large mountains slopes facing another mountain range at distances of $30-80$\,km) are associated with a detection rate well above the average. By splitting the detector into smaller sub-arrays of a few 10 000 km$^2$ each, deployed solely on favorable sites, an order-of-magnitude improvement in sensitivity could be reached with only a factor-of-3 increase in detector size, compared to the 60 000 km$^2$ simulation area. This is the envisioned GRAND setup. 

This neutrino sensitivity corresponds to a detection rate of 1 to 60 cosmogenic events per year. Besides, the angular resolution on the arrival directions, computed following \cite{Ardouin11}, could be as low as 0.05$^\circ$ for a 3~ns precision on the antenna trigger timing, opening the door for neutrino astronomy. 

\begin{figure*}
\centering
\includegraphics[width=5cm,clip]{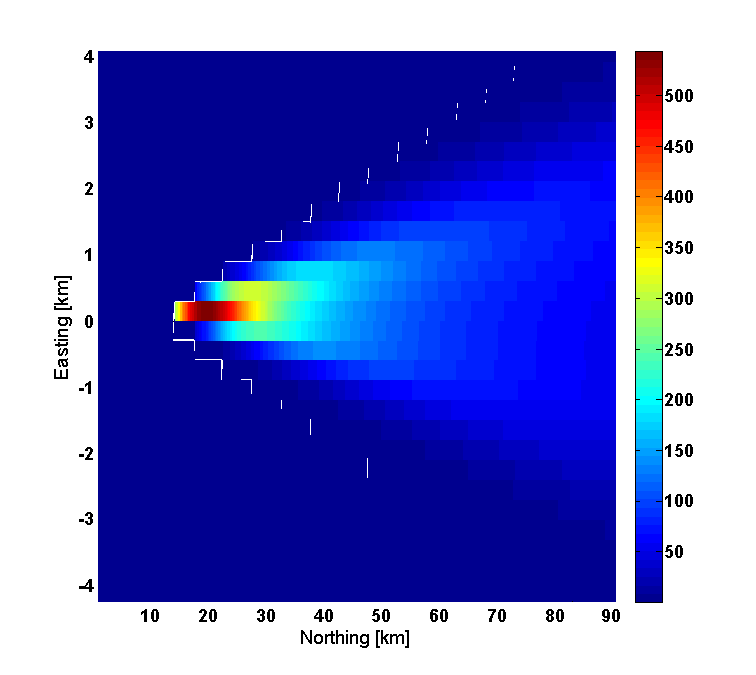} 
\includegraphics[width=5cm,clip]{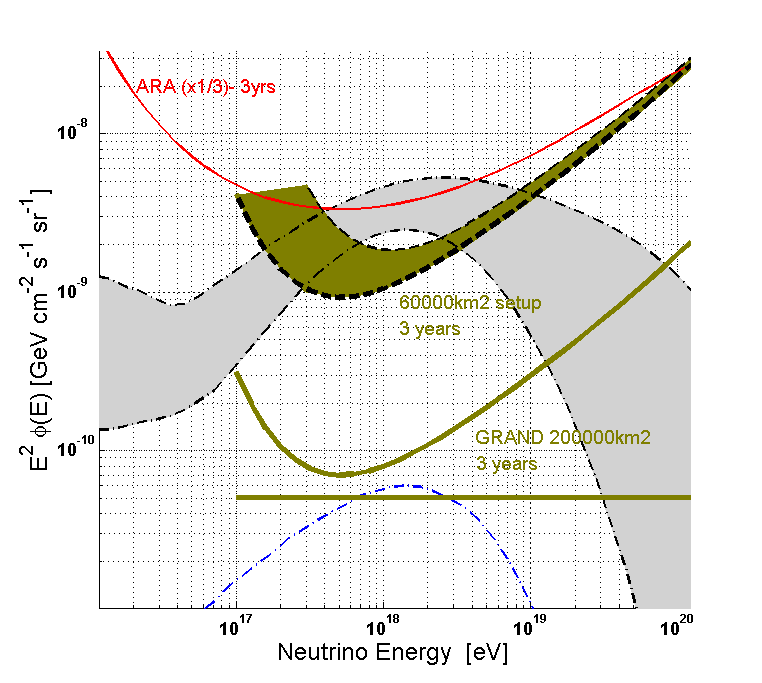} 
\caption{ {\it Left:} Expected radio footprint for a $5\cdot10^{17}$~eV horizontal shower induced by a tau decay at the origin. The color coding corresponds to the Efield maximum amplitude integrated over the 30-80MHz range (in $\mu$V/m). The sky background level is $\sim$15$\mu$V/m in this frequency range. Note the different x and y scales. {\it Right:} Differential sensitivity of the 60\,000 km$^2$ simulated setup (brown region, top limit: conservative, bottom: aggressive) and of the projected GRAND array (brown thick curve). The integral sensitivity limit for GRAND is shown as a thick line. We also show the expected limit for the projected final configuration of ARA~\cite{ARA2016} and theoretical estimates for cosmogenic neutrino fluxes \cite{KAO10}: the blue line stands for the most pessimistic fluxes, the gray-shaded region  to the ``reasonable'' parameter range. All curves are for single-flavor neutrino fluxes.}
\figlab{fig}     
\end{figure*}

\section{Background rejection}\seclab{bg}
A few tens of cosmogenic neutrinos per year are expected in GRAND. The rejection of events initiated by high-energy particles other than cosmic neutrinos should be manageable \cite{ICRC2015}. The event rates associated to terrestrial sources (human activities, thunderstorms, etc.) are difficult to evaluate, but an estimate can be derived from the results of the Tianshan Radio Experiment for Neutrino
Detection (TREND). TREND~\cite{Ardouin11} is an array of 50 self-triggered antennas deployed over a surface $\gtrsim1$\,km$^2$ in a populated valley of the Tianshan mountains, with antenna design and sensitivity similar to what is foreseen for GRAND. The observed rate of events triggering six selected TREND antennas separated by $\sim$800~m over a sample period of 120 live days was found to be around 1~day$^{-1}$, with two-thirds of them coming in bursts of events, mostly due to planes. Direct extrapolation from TREND results thus leads to an expected event rate of $\sim1$~Hz for GRAND for a trigger algorithm based on coincident triggers on neighbouring antennas and a rejection of events bursts. 

Amplitude patterns on the ground (emission beamed along the shower axis and signal enhancement on the Cherenkov ring \cite{Alvarez2012}), as well as wave polarization \cite{Aab14} are strong signatures of neutrino-initiated EAS that could provide efficient discrimination tools for the remaining background events. 

These options are being investigated within GRAND, through simulations and experimental work. In 2017 the GRANDproto project \cite{Gou15} will deploy a hybrid detector composed of 35 3-arm antennas (allowing for a complete measurement of the wave polarization) and 24 scintillators, that will cross-check the EAS nature of radio-events selected from a polarization signature compatible with EAS.

\section{GRAND development plan}\seclab{engineering}
Before considering the complete GRAND layout, several validation steps have to be considered. The first one will consist of establishing the autonomous radiodetection of very inclined EAS with high efficiency and excellent background rejection, with a dedicated setup of size $\sim 300$\,km$^2$. This array will be too small to perform a neutrino search, but cosmic rays should be detected above $10^{18}\,$eV. Their reconstructed properties (energy spectrum, composition) will enable us to validate this stage. The absence of events below the horizon will confirm our EAS identification strategy. A second array, 10 times larger, will allow to test the technological choices for the DAQ chain, trigger algorithm and data transfer. This will mark the start of GRAND data taking, foreseen in the mid-2020s.


%

\section{Conclusion}
The GRAND project aims at building the ultimate next-generation neutrino telescope. Preliminary simulations indicate that
 a sensitivity guaranteeing the detection of cosmogenic neutrinos is achievable. Work is ongoing to assess GRAND achievable scientific goals and the corresponding technical constraints. Background rejection strategies and technological options are being investigated.\\

\noindent\footnotesize{{\it Acknowledgements.} The GRAND and GRANDproto projects are supported by the Institut Lagrange de Paris, the France China Particle Physics Laboratory, the Natural Science Fundation of China (Nos.11135010, 11375209), the Chinese Ministry of Science and Technology and the S\~ao Paulo Research Foundation FAPESP (grant 2015/15735-1).}

\bibliography{ARENA}
\end{document}